\documentclass[11pt]{article}
\usepackage{graphicx}
\usepackage{amssymb}
\usepackage{amsmath}
\usepackage{pgfplots}
\usepackage{subfigure}
\usepackage{epstopdf}
\usepackage{subfigure, cite, hyperref, cleveref}
\usepackage{float}
\usepackage{collectbox}
\title{\  Non - Topological Solitons in a Non-minimally Coupled Scalar Field Induced Gravity 
Theory}
\author{{Daksh Lohiya\thanks{E--mail : dl116@cam.ac.uk}, and
Vartika Gupta\thanks{varg@physics.du.ac.in}}\\
       {\em *DAMTP, University of Cambridge}\\
       {Cambridge, CB3 0WA, UK}\\
       {\em \&}\\
       {\em Department of Physics and Astrophysics}\\
       {\em University of Delhi, Delhi 110007, India}
       }

\begin {document}
\baselineskip=2\baselineskip
\maketitle
\baselineskip= 15pt
\vskip 1cm
\centerline{\bf Abstract}
Properties of soliton stars that could be expected to naturally arise out of a first
order phase transition in non-minimally coupled scalar-field-induced 
gravity theories are investigated. Of particular interest are  
configurations, similar to Lee-Wick stars, with vanishing effective gravitational 
constant in the interiors.  

\vfil\eject

\section
{\bf A generalized scalar tensor theory}
\vspace{.5cm}

While there are no direct experiments to test gravity as described by
standard general relativity over scales less than (say) a fraction of a millimeter, 
it is now fairly well established 
that over galactic or larger scales, concordance of observations with general relativity 
is achieved only if we invoke a (theoretically unacceptably small) cosmological constant 
as well as a presence of a mysterious dark matter component. The {\it{un-natural}} values of
these parameters have motivated studies of dynamic dark energy models inspired by 
alternative gravity theories. Of particular interest are 
Scalar Tensor theories of gravity that have been studied right from the earlier Jordan 
\cite{jordan}, Brans-Dicke \cite{brans} 
and the Hoyle-Narlikar \cite{HN} versions in the 1960's to their more recent "avatars" and generalizations
\cite{many}. Most of these renditions are special cases of Horndeski's consolidated expression \cite{horn} 
for a general second order Scalar - Tensor theory in four dimensions. Variants of
these expressions have appeared in studies of dark energy and {\it{tracker field}} cosmological
models. These studies have included general (even non polynomial) expressions for an effective
potential for the scalar field as well as for the function of the scalar field that non-minimally 
couples to the Ricci scalar of the underlying spacetime. The kinetic term for the scalar
field could also come multiplied by an arbitrary function of the scalar field.  
A generalized Scalar - Tensor theory with a kinetic term $\approx w(\phi)\partial^\mu\phi\partial_\mu\phi$,
can, by a simple redefinition of the scalar field $\phi \longrightarrow \phi(\psi)$, be expressed 
in terms of the field $\psi$ having a canonical 
kinetic term $\propto \partial^\mu\psi\partial_\mu\psi$. In terms of $\psi$, the effective 
potential for $\psi$, as well as the coupling function of $\psi$ with the Ricci scalar would in general
be non-polynomial.  
The matter part of the action is chosen to be independent 
of the scalar field and with only a minimal coupling with the metric. This leads to the 
conservation of the stress tensor of the matter field and the geodesic motion for a 
pressureless fluid element of matter (dust). To have the construction indistinguishable
from general relativity, such renditions have to rely on some dynamical account for a rapid approach of the non-minimal
function of the scalar field to a constant value.\\

There have also been scalar - tensor models that hold the scalar field anchored to a constant value determined
by the minimum of the scalar field potential \cite{zee}. We consider a generalization of an 
interesting model proposed by Zee in which the action for the scalar tensor theory reads:
\begin{equation}
\label{1} 
S = \int d^4x\sqrt{-g}[U(\phi)R + \frac{1}{2} f(\phi)g^{\mu\nu} \partial_\mu\phi\partial_\nu\phi
- V(\phi) + L_m]
\end{equation}
Here $\phi$ is a scalar field, $V(\phi)$ its effective potential and $U(\phi)$ a non-minimal function that
describes a coupling of the scalar field with the Ricci scalar. $f(\phi)$ is yet another function multiplying
the kinetic term of the scalar field. $L_m$ is the Lagrangian for other matter fields.
The original version of the Scalar Tensor (Brans Dicke) theory used 
$U(\phi) = \phi = \omega /f(\phi)$, however, one can explore more general functions. \\
If the scalar field gets dynamically 
locked at the minimum of the scalar potential $V(\phi)$ at $\phi = \phi_o$, it is easy to see that the action,
at $\phi = \phi_o$, is indistinguishable from the canonical Einstein - Hilbert action in general 
relativity with the identification 
\begin{equation}
16\pi G_N = [U(\phi_o)]^{-1}
\end{equation}
and with an effective cosmological constant determined by 
$V(\phi_o)$.
The equations of motion that follow from the variation of the action Eqn(1) with respect to the metric 
and the scalar field are:
\begin{equation}
U(\phi)G^{\mu\nu} = -\frac{1}{2}[T_m^{\mu\nu} + T_\phi^{\mu\nu} + 2U^{;\mu\nu} - 2g^{\mu\nu}U^{;\eta}_{;\eta}]
\end{equation}

\begin{equation}
f(\phi)\Box\phi+\frac{1}{2}f'(\phi)\partial_\beta\phi\partial^\beta\phi+ V'(\phi) - U' R - {\partial L_m\over{\partial\phi}} = 0
\end{equation}

Here $T_m^{\mu\nu}$ is the energy momentum tensor of the rest of the matter fields, assumed to be independent 
of $\phi$, and 
\begin{equation}
 T_\phi^{\mu\nu} = f(\phi)\partial^\mu\phi\partial^\nu\phi 
 - g^{\mu\nu}[\frac{1}{2}f(\phi)\partial^\eta\phi\partial_\eta\phi - V(\phi)]
\end{equation}
It is straightforward to demonstrate that with $L_m$ chosen to be independent of $\phi$, 
the rest of the matter field satisfies
equivalence principle expressed by way of the vanishing of the covariant divergence of the its stress energy 
tensor:  $T^{\mu\nu}_{m;\nu} = 0$. This follows from the Bianchi identity as well as the equation of motion 
of $\phi$ Eq(4). On the other hand, if the matter field has $\phi$ dependence, we get:
\begin{equation}
T^{\mu\nu}_{m;\nu} = \partial^\mu\phi\left[ \partial_\nu\left(\frac{\delta L_m}{\delta \partial_\nu\phi}\right) -
 \frac{\delta L_m}{\delta \phi}\right]
\end{equation}
However, if the scalar field is dynamically held to a constant value in a region, the vanishing of covariant 
divergence of the matter stress tensor $T_m^{\mu\nu}$ in that region is again assured. \\

There have been extensions of the action described by Eq(1) in which one subtracts away a surface terms given
identically by the non-minimal function convoluted with the trace of the second fundamental form at the boundary
$\partial M$ of the manifold $M$:
\begin{equation}
 S = \int_M d^4x\sqrt{-g}[U(\phi)R + \frac{1}{2}f(\phi)g^{\mu\nu} \partial_\mu\phi\partial_\nu\phi
- V(\phi) + L_m] - \int_{\partial M} d\Sigma^\mu K_\mu U(\phi)
\end{equation}
In a region where $\phi$ is constant, standard general relativity again follows. The surface term merely takes 
away the second derivative terms of $U(\phi)$ from Eq(3). For a general (varying) non - minimal coupling term,
such a subtraction is essential to define a consistent {\it{quasi conformal mass}} in the theory 
\cite{york,mann,bose}. The addition of such a surface term again leads to a violation of the equivalence principle,
giving instead:
\begin{equation}
 T^{\mu\nu}_{m;\nu} = -2U_\nu R^{\mu\nu} = -2U'(\phi)\partial_\nu\phi R^{\mu\nu}
\end{equation}
This is even if the matter part is independent of $\phi$. However again, in a region where $\phi$ is 
dynamically anchored to a constant value, the vanishing of the covariant divergence is again assured. \\

Zee used $U(\phi) = \frac{1}{2}\epsilon\phi^2$, $f(\phi) = 1$, and a scalar potential $V(\phi)$ having 
a vanishing minimum at $\phi = \phi_o$:
\begin{equation}
 V_{Zee}(\phi) = \frac{\lambda}{8}(\phi^2 - \phi_o^2)^2
\end{equation}
and demonstrated that with the scalar field $\phi$ anchored at the fixed 
minimum of the potential $\phi_o$, the theory is indistinguishable from general relativity at low energies. 
Any perturbations around the minimum of the potential are expressible in terms of extremely unstable 
scalar field excitations that quickly decay into gravitons. The mass of the unstable scalar particle
is approximately:
\begin{equation}
m_{eff}^2 = \frac{V''}{1 + \frac{3(U')^2}{U}}\arrowvert_{\phi = \phi_o}
\end{equation}
For $U(\phi) = \epsilon\phi^2/2$ used by Zee, this gives:
\begin{equation}
m_{eff}^2 = \frac{V''(\phi_o)}{1 + 6\epsilon}
\end{equation}
%In this article we entertain the possibility of the potential $V(\phi)$ having two almost 
%degenerate minima at say $\phi = 0$ and $\phi = \phi_o$, with the non-minimal 
%coupling becoming arbitrarily large at $\phi = 0$. Consider a model described by 
%Eq.[1] with a scalar potential
%\begin{equation}
%V(\phi) = \frac{1}{2}m_{\phi}^2\phi^2[1 - \frac{\phi}{\phi_o}]^2 
%+ B[4(1 - \frac{\phi}{\phi_o})^3 - 3(1 - \frac{\phi}{\phi_o})^4]
%\end{equation}
In this article we propose considering a non-minimal 
coupling function $U(\phi)$ that can acquire an arbitrarily large value at $\phi = 0$, assured for 
example by (but not restricted to) a simple choice:
\begin{equation}
 U(\phi) = \frac{M^4}{\phi^2 + \epsilon^2}
\end{equation}
with $\epsilon << \phi_o$ and very small
$$
M_{Plank}^2 \equiv  \frac{M^4}{\phi_o^2 + \epsilon^2} \approx \frac{M^4}{\phi_o^2}
$$
The choice of the potential $V(\phi)$ in this article would be one that has a zero that
is a local minima at $\phi = \phi_o$, and has another minima at $\phi = 0; V(0) > 0$. With the 
choice of the non-minimal coupling function $U(\phi)$ given by Eq(12), we show below that the 
dynamics of the scalar field 
is determined by an effective potential $W(\phi)$ that has a minimum at both $\phi = \phi_o$
as well as $\phi = 0$. We shall require and choose the profile of potential $V(\phi)$,
and the function $U(\phi)$,
in the interval $(0,\phi_o)$ to be such that 
the scalar field dynamics supports non-trivial configurations having the field $\phi$
locked to a vanishing value inside a sphere and transiting to $\phi_o$ outside across
a thin wall.\\

While a non-minimally coupled Scalar Tensor theory, in what is referred to as 
the {\it{``Jordan frame''}}, can be related to a minimally coupled scalar field
theory, by a conformal transformation to an {\it{``Einstein frame''}}, such a 
transformation is excluded for a divergent $U(\phi)$. Further, in what follows, a ``Higgs'' coupling of
the scalar field to a Fermion field would give distinctive mass for the fermions 
at multiple minima of the effective potential and different effective gravitational ``constants'' 
in regions confining fermions. These constructs constrains us to work in the Jordan
frame - and look for solutions having the scalar field locally locked to a minima of the 
effective potential.\\
%Consider a model described by 
%$Eq.[1] with a scalar potential
%$\begin{equation}
%$V(\phi) = \frac{1}{2}m_{\phi}^2\phi^2[1 - \frac{\phi}{\phi_o}]^2 
%$+ B[4(1 - \frac{\phi}{\phi_o})^3 - 3(1 - \frac{\phi}{\phi_o})^4]
%$\end{equation}

For the matter part, we consider two components: (a) a part that is independent of the 
scalar field, described by a term $L_{w}$ in the action, and (b) a fermion field having a $\phi$ 
- dependent Higgs coupling:
\begin{equation}
L_m \equiv L_\psi + L_{\psi,\phi} + L_w 
\equiv {1\over 2}[\bar\psi\overleftarrow{D}_\mu\gamma^\mu\psi
- \bar\psi\gamma^\mu\overrightarrow{D}_\mu\psi]
- m_f(\phi)\bar\psi\psi + L_w
\end{equation}
with $m_f$ the fermion mass parameter that varies from $0$ to $m_{ext}$ as $\phi$ 
goes from $0$ to $\phi_o$.
$$
m_f(\phi) = m_{ext}{\phi\over \phi_o}
$$
Here $D_\mu$ is the spin covariant derivative: 
%[see eg. \cite{__}]
\begin{equation}
 \overrightarrow{D}_\mu\psi = (\partial_\mu + \Gamma_\mu)\psi ; ~~~~~
\bar\psi\overleftarrow{D}_\mu = (\partial_\mu\bar\psi - \bar\psi\Gamma_\mu)
\end{equation}
$\Gamma_\mu$ are the spin connection [Fock - Ivanenko] coefficients
defined by \cite{pagels}:
\begin{equation}
D_\nu\gamma_\mu \equiv \partial_\nu\gamma_\mu 
- \Gamma^\alpha_{\mu\nu}\gamma_\alpha + [\Gamma_\nu,\gamma_\mu] = 0
\end{equation}

Features of such generalized Scalar Tensor theories, along with the expression for a conserved 
energy momentum (pseudo-) vector for such an 
action have been studied in detail in \cite{bose} and have been summarized in the Appendix.
For the purpose of 
the present article, it would suffice to note two trivial solutions: (a) at $\phi = 0$, the action simply 
describes a massless 
fermion and matter fields in flat spacetime while (b) at $\phi = \phi_o$, the action is indistinguishable 
from that of the Einstein action for matter fields as well as a massive fermion field (mass $m_{ext}$).
Besides these
trivial solutions, with the fermions number conserved, there would also exist non-trivial solutions in which 
a given number of massless fermions, that do not have enough energy to be on shell at $\phi = \phi_o$,
are trapped inside a region having $\phi = 0$ in the interior - with $\phi$ making a sharp transition 
across the boundary to the exterior having $\phi = \phi_o$.
For a given conserved number of massless fermions that are on shell at $\phi = 0$, 
it is straightforward to see that the model has the same features as that of a 
Lee-Wick model \cite{lee}. Properties of such non-trivial, non - topological soliton solutions 
are explored in section (2). However, we feel compelled to include our motivation in the 
next subsection:

\subsection{Why bother with non - minimal coupling}

Over the last more than thirty years there has been a consensus that the universe has been
through a series of first order phase transitions. Renditions of these transitions are schematically
and typically described in terms of the dynamics of a (minimally coupled) scalar field having a 
potential with non-degenerate minima. At sufficiently high temperatures, the effective potential 
has a unique minimum corresponding to say a phase ``A''. As the temperature drops to some 
critical temperature $T_c$, 
the phase ``A'' could co-exist in pressure and chemical equilibrium with another phase ``B'' that 
corresponds to the development of a second minimum of the effective potential. Bubbles of phase ``B''  
nucleate rapidly once the temperature cools sufficiently below $T_c$. Once nucleation of phase ``B'' 
bubbles commences, the latent heat (viz.: the difference between the ground state energies corresponding to 
the the two minimum of the effective potential) that is released, re heats the universe to 
$T_c$ - suppressing further formation of phase
``B'' bubbles. As the universe expands, the temperature is maintained at $T_c$ by the liberation of latent
heat as the volume of phase ``B'' grows with the shrinking of that of phase ``A'' \cite{kaj}. The scalar field 
could further have a Higgs coupling with fermions that are massless and therefore expected 
to be copiously produced throughout 
in the phase ``A'' region, but which have a sufficiently large mass in phase ``B''. As a result,  
percolation of bubbles of phase ``B'' throughout the volume of the universe would leave out pockets 
of phase ``A'' where fermions are trapped. These regions would survive as non - topological 
solitons.\\

Variations of this scenario have been used to account for the formation and existence of such 
solitons in many studies. Using phases ``A'' and ``B'' as the unconfined and confined 
phase of a quark gluon plasma, elaborate accounts have been reported for the formation of baryons 
as well as of ``$Lee - Wick$'' and ``$Lee - Pang$'' solitons. In this case the scalar field potential
parameters are chosen to endow the solitons with a ``$bag - pressure$'' $B = (100~MeV)^4$ and a 
``$surface- tension$'' $S = (30~GeV)^3/6$ that are characteristic energies used in hadron
spectroscopy. In a different model, with an appropriate and similar Higgs
structure, one can have two chirally degenerate ground states such that in one ground 
state (phase ``A'') the left handed Majorana neutrinos are massive but the right handed ones are 
massless, while in the other ground state (phase ``B''), the left handed neutrinos are massless
but the right handed ones are massive. In this case, the surface tension of the wall separating
the two phases was chosen to be $S \leq (1.93~TeV)^3$ to account for trapped right handed 
neutrino balls \cite{many1}.\\

A problem with the above schemes arises with the phase ``A'' having or acquiring a non - vanishing, 
positive minimum of the effective potential at high temperatures. This would endow the spacetime 
with an effective cosmological constant, resulting in exponential expansion of the phase ``A''. 
An account of a desirable percolation of phase ``B'' is not forthcoming in such an event. This was 
also the primary issue that led to the abandoning of the {\it{old inflationary cosmology model}}.\\

On the other hand, a non - minimally coupled scalar field has some very desirable features. In the 
presence of a cosmological constant, the scalar field develops a condensate that identically 
cancels out: not only the cosmological constant but the effective gravitational constant as well, with 
the expansion scale factor of the universe quickly approaching a linear (instead of the 
exponential) expansion in time \cite{fordetc}. 
It is this feature that is explored in conjunction with attributes of an effective potential with two (in general) 
non - degenerate minima in this article. To make matters simple, the feature of dynamic divergence 
of the non - minimal coupling in the presence of an effective cosmological constant in models 
studied by Ford, Dolgov et al., is, by a redefinition of the scalar field, equivalently 
incorporated in Eqn(1 \& 12) by vanishingly small $\epsilon$ that results in $U(\phi)$, given by Eqn(12), to become arbitrarily large
as $\phi \longrightarrow 0$ for $\epsilon \ll M^2$. A hot universe with the scalar field at the minimum
of the effective potential at $\phi = 0$, would quickly approach a linear expansion. 
Percolation of the phase ``B'' bubbles 
would no longer be a problem and any of the phase transition  scenarios could be appropriately accounted for.\\

The non - topological solitons that would arise from trapped fermions at $\phi = 0$ domains 
have, by construction, very simple properties. Their interior has a vanishing effective 
{\it gravitational constant}, the exterior the canonical gravitational constant, and the metric of 
the exterior is just the Schwarschild metric. The
total energy of these solitons - specially the gravitational part, can be exactly determined.

\section
{\bf Non-Topological Solitons}
\vspace{.5cm}
We explore the possibility of classical solutions to Eqns(3 \& 4) for a fixed number of massless fermions 
trapped inside a spherically symmetric static region as the scalar
field makes a transition from $\phi = 0$ in the interior of the region to $\phi = \phi_o$ across a thin 
wall containing the region. Using 
isotropic coordinates (that would be appropriate to describe the solution for a size greater than the 
Schwarzschild bound), the metric is given by eqns(A.30) and (A.32) outside and inside the boundary 
respectively. 
Denoting the fermion density by $S_f$, eqn(4) reads:
$$
f(\phi)\Box\phi + {1\over 2}f'(\phi)\partial_\beta\phi\partial^\beta\phi  
 + V'(\phi) - U' R - {\partial m_f\over {\partial \phi}}S_f  
= 0 \eqno{(16)}
$$
As shown in \cite{lee}, $S_f$ is described in terms of the chemical potential and temperature of the 
fermion gas. The trace of eqn(3) gives:
$$
U(\phi)R = -(3U''(\phi) +{f(\phi)\over 2})\phi^{,\alpha}\phi_{,\alpha}
+ 2V(\phi) + {1\over 2}m_fS_f - 3U'(\phi)\Box\phi \eqno{(17)} 
$$
Substituting in eqn(16) gives: 
$$ 
\left(f(\phi) + 3{U'^2\over U}\right)\Box\phi  
+ {1\over 2} \phi^{,\alpha}\phi_{,\alpha}\left[{U'\over U}\left(6U''(\phi) + f(\phi)\right) + f'(\phi)\right]
$$
$$
+ V' - m_f'S_f - {U'\over U}\left(2V + {1\over 2}m_fS_f\right) =0  \eqno{(18)}
$$
While this system could well have non - trivial solutions, to demonstrate the existence of soliton solutions
we restrict our choice of the function $f(\phi)$ so that the ``kinetic'' term in Eqn(18) vanishes. We look for 
$f(\phi)$ that solves:
$$
U(\phi)f'(\phi) + U'(\phi)f(\phi) + 6U'U'' = 0
$$
$$
\Rightarrow f(\phi) = -{(3U')^2\over U} + {\omega\over U} \eqno{(19)}
$$
This is a generalization of the choice for $f(\phi)$ made by Brans and Dicke and simplifies Eqn(18) to:
$$ 
\Box\phi + U\left(V' - m_f'S_f\right) - U'\left(2V + {1\over 2}m_fS_f\right)  = 0 \eqno{(20)}
$$
where we have absorbed the factor of $\omega$ in a redefinition of $V(\phi)$. 
In a {\it {thin wall approximation}}, with the fermions being massless at $\phi = 0$ in the 
interior, and the fermion density vanishing in the exterior, one could neglect $S_f$ at
the boundary. Further,
as described in the Appendix Eqns(A.30 \& A.32), with a flat metric in 
the interior connecting continuously with the exterior metric, for a static spherically symmetric 
configuration, this eqn(20) reduces to:
$$
{d^2\phi\over dr^2} + {2\over r}{d\phi\over dr} = {dW\over d\phi} \eqno{(21)}
$$
where
$$
{dW\over d\phi} \equiv  UV' - U'2V \eqno{(22)}
$$
For the choice of the non - minimal function Eqn(12), a choice of $V$ having, in general even non - 
degenerate, minima at $\phi = 0$ and $\phi = \phi_o$, would ensure that $W(\phi)$ also  has minima at 
these very values, and further the profile of $V(\phi)$ could be chosen to support non trivial solutions 
to Eqn(21) with $\phi = 0$ in the interior and $\phi = \phi_o$ (the true ground state) in the exterior. 
These configurations would be similar to the Lee-Wick stars \cite{lee}.\\

This behavior sets a stage for a scenario in which at High temperatures the effective potential of the scalar
field could have one true ground state at $\phi = 0$, and at which temperature copious amount of 
massless fermions would be produced.
As the universe cools, bubbles of true vacuum phase $(\phi = \phi_o)$ could be formed. At temperatures 
much below the fermion effective mass $m_f$, massless fermions would get constricted to the the interior of non - 
topological solitons: regions with $\phi = 0$. 

The energetics and stability of these solitons follow from similar analysis of soliton stars in flat
spacetime \cite{cott, akg} along with the exact expression for the gravitational energy given by Eq(A.33).
This is used in the next section.

\vspace{.5cm}
\section
{\bf Soliton Stars - cold and hot}
\vspace{.5cm}

As described in the Appendix, in asymptotically Minkowskian coordinates, the conserved energy 
contained inside the radial parameter $\rho = \rho_o$, for a 
spherically symmetric, static system, in which the non - minimally coupled scalar field is anchored
to the true ground state $\phi = \phi_o$ outside $\rho = \rho_o$ is simply and exactly given by Eq.(A.33).  
The mass parameter $M$ is the conserved mass at infinity. In other words, if a total energy $M$ is slowly
lowered from infinity in a spherically symmetric manner to  $\rho_o$, the isotropic radial parameter, 
and thereafter this energy gets re distributed as a thin wall non-topological soliton with $\phi = 0$
in the interior $\rho < \rho_o \Longleftrightarrow r < r_o$, the total energy is given by Eq(A.33) 
(here $r$
is the Schwarzschild radial parameter). 
This justifies the ansatz of using the
flat space expression for the energy for $M$ with the gravitational correction given by 
$$
E_G =  - {M^2G\over {2c^2 \rho_o}}
= - {Mm\over {(r_o - m + \sqrt{r_o^2 - 2mr_o})}}; ~~{\rm where}~~m \equiv {GM\over c^2} \eqno{(23)}
$$
This incidentally tallies with the work done to bring a total mass $M$ to the radius $\rho = \rho_o$ 
in Newtonian gravity, or the work done to assemble a shell of mass $M$ of radius $\rho_o$.\\

  We now recall expressions for energy of a Lee-Wick soliton in flat spacetime \cite{cott, akg}. 
The thermodynamic potential $\Omega$, the free energy $F$ and the partition function $Z$ for a gas of
fermions are related by:
$$
Z = e^{-\beta\Omega};~~~~ F = \Omega + \mu N \eqno{(24)}
$$
Where $\mu$ is the chemical potential and $N$ the conserved fermion number. The internal energy of the 
gas and the thermodynamic potential are:
$$
E = {\partial\over {\partial\beta}}(\beta F)\eqno{(25)}
$$
$$
\Omega = - {2v\over \beta}\int{d^3k\over {(2\pi)^3}}[{\rm ln}(1 + e^{-\beta(E_k - \mu}) 
+ {\rm ln}(1 + e^{-\beta(E_k + \mu})] \equiv \bar\Omega v \eqno{(26)}
$$
The factor of $2$ is on account of two spin states of a fermion and would be absent in case of
massless neutrinos. This may be replaced by $n_f$: the number of massless neutrino flavours. 
The fermion number $N$ and number density $n$ are:
$$
N = - \left[{\partial\Omega\over \partial\mu}\right] = - \left[{\partial\bar\Omega\over \partial\mu}\right]v
\equiv nv\eqno{(27)} 
$$
For temperatures much less than the effective mass of the scalar field excitations, the contribution of
scalar particles to the thermodynamic potential is suppressed. The contribution would come only from the 
fermion gas.  The transition of the scalar field across 
a thin wall would contribute to a surface energy $E_s$, while the value of the scalar field at $\phi = 0$
contributes to a volume energy $E_v$:
$$
E_s = 4\pi r^2 S ~~~~~~~ E_v = {4\over 3}\pi r^3 V(0)  \eqno{(28)}
$$
where we treat the surface tension $S$ and the pressure $V(0)$ as free parameters and $r$ is the 
radius of the shell. Ignoring gravitational 
effects in the first approximation, the total free energy is 
given in terms of the chemical potential and the thermodynamic potential as:
$$
F = \mu nv + \bar\Omega v + E_s + E_v = \mu N + \bar\Omega v + E_s + E_v \eqno{(29)}
$$
At zero temperature:
$$
\bar\Omega = -{\mu^4\over {12\pi^2}}; ~~ n = {\mu^3\over {3\pi^2}}; ~~N = {4\over 9\pi}\mu^3r^3 
\Rightarrow 
F = \left({\mu^4\over {4\pi^2}} + V(0)\right){4\over 3}\pi r^3 + 4\pi r^2S \eqno{(30)}
$$
For a fixed conserved $N$ number of fermions, using Eq(27) to eliminate $\mu$, gives:
$$
%F = \alpha {N^{4/3}\over r} +  V(0)\right){4\over 3}\pi r^3 + 4\pi r^2S; ~~~~{\rm where}~~ 
\alpha = {\pi^{1/3}\over 2}\left({3\over 2}\right)^{5/3} \eqno{(31)}
$$
We extremize the free energy in Eq(29) under the change of volume (radius) for fixed $N$ using the method
of Lagrange multiplier. Variation of free
energy with respect to the volume being proportional to external pressure, the extremization is equivalent
to putting the external pressure to zero for fixed $N$, and gives:
$$
\bar\Omega + V(0) + {2S\over r} = 0  \eqno{(32)}
$$
From the expression for $\bar\Omega$, this gives:
$$
8\pi S r^2 = 4\pi^3r^3\left({\mu^4\over {12\pi^4}} - {V(0)\over \pi^2}\right)
= v\left({\mu^4\over {4\pi^2}}\right) - 4\pi r^3V(0) \eqno{(33)}
$$
substituting in the expression for free energy Eq(30), this gives the minimum:
$$
M \equiv F_{min} = 4V(0)v + 12\pi r^2S \eqno{(34)}
$$
For $V(0) = V(\phi_o) = 0$ (degenerate vacuum), that defines the Lee - Pang model, eliminating $\mu$, one gets:
$$
r_{min} = \left({\alpha\over 8\pi S}\right)^{1/3}N^{4/9}; ~~~~~~~ M = {3\over 2}\alpha^{2/3}(8\pi S)^{1/3}N^{8/9}
\eqno{(35)}
$$
Stability against break up into smaller units is ensured by the exponent of $N$ being less than unity. 
For large $N$, gravitational effects would have to be considered. These follow from the expressions
Eqns (23); (31):
$$
E(r) = \bar E - {{\bar E m_{\bar E}}\over {r - m_{\bar E} + \sqrt{r^2 - 2m_{\bar E}r}}}
$$
$$~{\rm where} ~~
m_{\bar E} \equiv {G\bar E\over c^2};~~ \bar E \equiv \alpha {N^{4/3}\over r} + 4\pi r^2S \eqno{(36)}
$$
On the other hand if the vacuum is non - degenerate, with $V(\phi_o) \equiv B > 0$, 
%%%%%%%
$$
\bar E = \alpha \frac{N^{4/3}}{r} + {4\over 3}\pi Br^3 + 4\pi r^2S \eqno{(37)}
$$
To compare the results with standard results of the ``Lee - Pang'' model, we assume that the non - minimally 
coupled scalar field field theory with an arbitrarily large $U(0)$ presented here describes the dynamics of 
the QGP phase transition. Further, for $B$ and $S$ we assume the values 
$B_o = (100~MeV)^4; ~ S_o = (30~GeV)^3/6$ typically used 
in Hadron spectroscopy. This reduces the above expressions for the energy (in grams) to:
$$
E_{dgen} \approx 5.5 \times 10^{-23}N^{8/9}E_1(x)\left[{1 - 2\epsilon_1N^{4/9}E_1(x)/x + 
\sqrt{1 - 2\epsilon_1N^{4/9}E_1(x)/x}}\over {1 - \epsilon_1N^{4/9}E_1(x)/x + 
\sqrt{1 - 2\epsilon_1N^{4/9}E_1(x)/x}}\right] \eqno{(38)}
$$
for the degenerate case. A dividing factor of $M_\odot\approx 2\times 10^{33}$ would give the energy in Solar Mass. Here $r \equiv r_{min}x$ and
$$
E_1(x) = {2\over x} + x^2; ~~ r \approx 4.59\times 10^{-16}N^{4/9}x~{\rm cm}; 
~~{\rm and}~~ \epsilon_1 \approx 9.35 \times 10^{-36}\eqno{(39)}
$$
%-------------figure for zero temp-------------------------------------
%------------------------------------------------------------------
\begin{figure}%[H]
\centering
\mbox{\subfigure[]{\includegraphics[width=67mm, height=58mm]{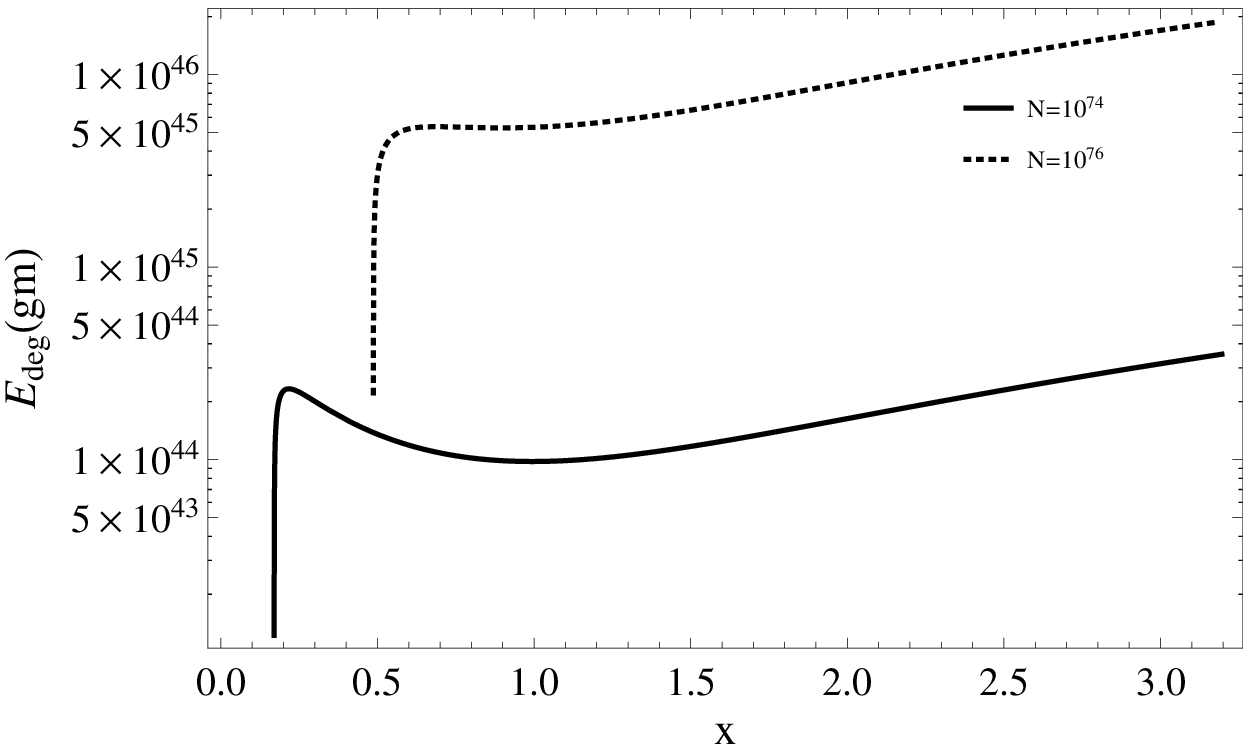}}
\subfigure[]{\includegraphics[width=67mm, height=58mm]{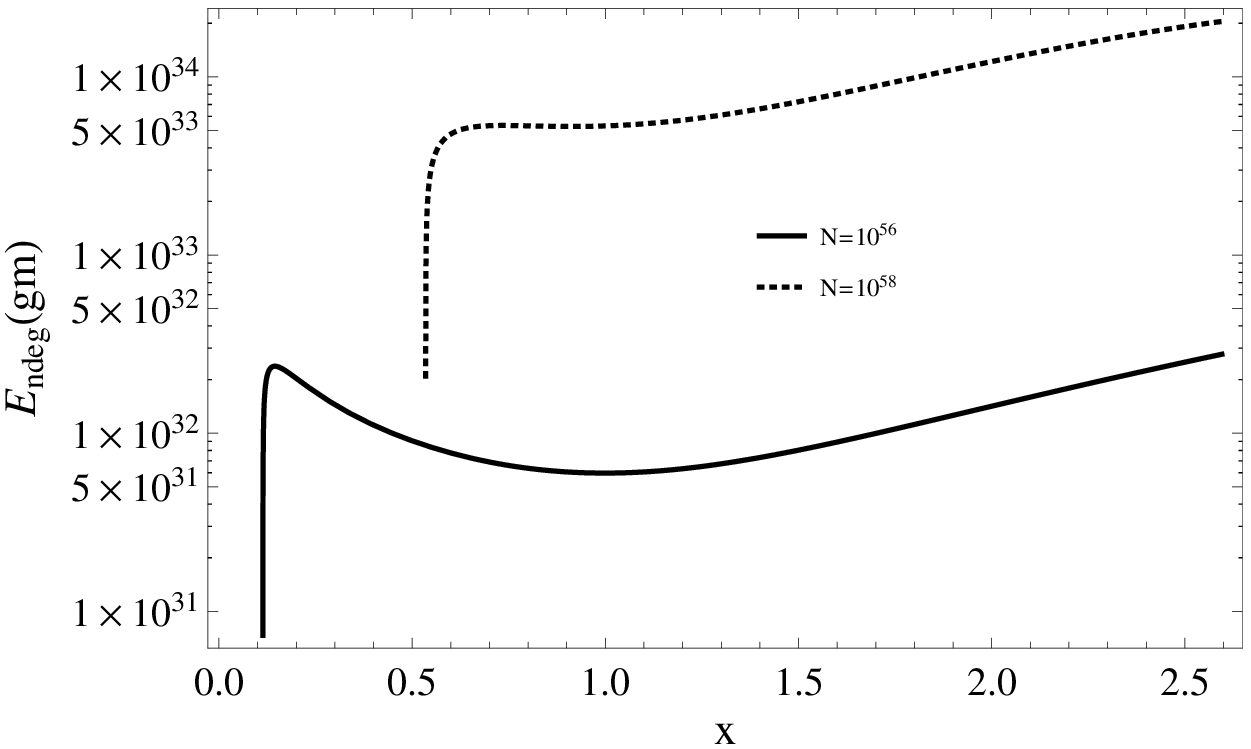}}}
\label{fig2}
\caption{ Left panel shows the variation of free energy at zero temperature for degenerate vacuum with $x$, for two different values of N . Right panel shows the same for non-degenerate vacuum. }\label{fig:b0}
\end{figure}
The above values are given for $S = S_o = (30 {\rm GeV})^3/6$. For a general value there would be an
overall multiplicative factor of  $(S/S_o)^{1/3}$ in Eq(38) aside from factors of $(S_o/S)^{1/3}$ in the 
expression for $r$ and $(S/S_o)^{2/3}$ for $\epsilon_1$ in Eq(39).
For the non-degenerate case, the energy (in grams) is
$$
E_{ndgen} \approx 4.5 \times 10^{-25}NE_2(x)\left[{1 - 2\epsilon_2N^{2/3}E_2(x)/x + 
\sqrt{1 - 2\epsilon_2N^{2/3}E_2(x)/x}}\over {1 - \epsilon_2N^{2/3}E_2(x)/x + 
\sqrt{1 - 2\epsilon_2N^{2/3}E_2(x)/x}}\right] \eqno{(40)}
$$
where $r \equiv (\alpha/4\pi B)^{1/4}N^{1/3}x$ and
$$
E_2(x) = {1\over x} + \bar\epsilon x^2 + {x^3\over 3};~ 
\bar\epsilon \approx {7.74\times 10^6\over {N^{1/3}}}; ~
\epsilon_2 \approx 3\times 10^{-40}; ~r \approx 10^{-13}N^{1/3}x~{\rm cm} \eqno{(41)}
$$
For general values of the parameters, there would be an
overall multiplicative factor of  $(B/B_o)^{1/4}$ in Eq[40] aside from factors of $(B_o/B)^{1/4}$ in the 
expression for $r$, $(B/B_o)^{1/2}$ for $\epsilon_2$ and $(S/S_o)(B_o/B)^{3/4}$ for $\bar\epsilon$. \\

%\begin{figure}[H]
%\centering
%\subfigure[]{\includegraphics[width=60mm, height=54mm]{56n58ndeg.jpeg}\label{fig2}}
%\caption{  Plot of $E_{ndeg}$ with two different values of N . }\label{fig:b0}
%\end{figure}
The choice of parameters $S$ and $B$ determines the size of the solitons and the number  of fermion determines the stability: with gravity duly incorporated. 
To compare with the results in \cite{cott, akg}, the values used in hadron spectroscopy are used in 
Figure(1) to find a limit of $N \geq 10^{76};~~ r_c \approx 2.91 ~$light year;$~
M_c \approx 2.957*10^{12}~M_{\odot}$ for the degenerate case. For the non-degenerate case, the limits are
$N \geq 10^{58};~~ r_c \approx 24.69$Km$;~M_c \approx 2.65~M_{\odot}$. For a neutrino ball, considering two
flavours of trapped right handed Majorana neutrinos, and using $S = (1.9 {\rm TeV})^3$, gives the limit
of $N \geq 10^{72};~~ r_c \approx 3.99*10^{12}$~m$;~
M_c \approx 5.55*10^{9}~M_{\odot}$ for the degenerate case. One could have similar limits for the non-degenerate case by 
chosing the parameter $B$.

%  Vartika, just determine these masses in solar mass units, enclose just a curve for N some two orders of
%magnitude below the development of the saddle point and at the saddle point with the x range 
% between 0.1 and say 2.5. This way the minimum at x = 1 would show well - rather than being on just one side.
%[COMMENT ON HOW THESE 
%COMPARE WITH GALAXY OR NEUTRON STAR SIZES OR SAGITARIUS-A IN SIZE AND MASS BY CHANGING THE RATIO OF 
%$S^{1/3} AND B^{1/4}$]\\

At non-vanishing temperatures, the above analysis significantly changes when finite temperature field theory methods are applied. We determine the equilibrium by minimizing the free energy of the
system. Assuming that the contribution of the scalar field to the free energy density is given by the 
expression in the absence of gravitation as \cite{cott}:
$$
F(\phi) = V(\phi) + {1\over \beta}\int{{d^3k}\over {(2\pi)^3}}{\rm ln}[1 - exp(-\beta\sqrt{k^2 + M_\phi^2})];
$$
$$
{\rm with}~ \beta \equiv {1\over {k_BT}}; ~~{\rm and} M_\phi^2 \equiv V''(\phi)\eqno{(42)}
$$
For low enough temperatures for which $k_BT$ is much less than both $V''(\phi_o)$ and $V''(0)$ as well 
as the effective mass of the fermions at $\phi = \phi_o$; 
the temperature dependent contribution from the scalar field to the free energy is exponentially small as 
$\beta M_\phi >> 1$ both inside as well as outside the soliton. The scalar field free energy density is thus
unaltered from its value of zero outside the soliton to $V(0) = B$ inside. For the same reason, the 
fermion free field energy is unaltered outside the soliton. However, inside the soliton, the fermions 
being massless, copious amounts of fermions would be produced at any temperature. The expression for free
energy is determined from Eqns[24 to 27]. 
$$
\bar\Omega = -{1\over \beta^4}\left[{{(\mu\beta)^4}\over {12\pi^2}} + {(\mu\beta)^2\over 6} + 
{7\pi^2\over 180}\right] \eqno{(43)}
$$
$$
n = {\mu^3\over 3\pi^2} + {\mu\over 3\beta^2}; ~~ \Rightarrow N = {4\pi r^3\over 9\beta^3}
\left[\mu\beta + {(\mu\beta)^3\over \pi^2}\right] \eqno{(44)}
$$
$$
F = {1\over \beta^4}{4\pi\over 3}r^3\left[{{(\mu\beta)^4}\over {4\pi^2}} + {(\mu\beta)^2\over 6} 
+ B\beta^4 - {7\pi^2\over 180}\right] +4\pi r^2S \eqno{(45)}
$$
Minimizing this $F$ for fixed $N$, that amounts to having zero external pressure, gives Eqn(32). It is convenient 
to determine the mass of the soliton from the free energy that follows from Eqn(25):
$$
E = F + \beta{\partial\Omega\over \partial\beta} = \left[{\mu^2\beta^2\over 2} + {\mu^4\beta^4\over {4\pi^2}}
+ \bar\alpha + {7\pi^2\over 45}\right]{v\over \beta^4} + 4\pi r^2S \eqno{(46)}
$$
where $\bar\alpha \equiv B\beta^4 - 7\pi^2/180$. Thus again we have a term proportional to the volume and 
a surface area dependent term. The volume dependent term defines the energy density:
$$
\varrho = \left[{\mu^2\beta^2\over 2} + {\mu^4\beta^4\over {4\pi^2}}
+ \bar\alpha + {7\pi^2\over 45}\right]{1\over \beta^4}
$$
Inclusion of gravitational correction is straightforward and follows simply by replacing $\bar E$  in Eqn(36)
by the above expression for $E$ in Eqn(46). \\

Stability of the soliton follows from the minimization of the free 
energy. For $(\mu\beta)$ small, to the order of $(\mu\beta)^2$, without the gravitational corrections 
the free energy can be expressed in terms of the fermion number using Eqn(44). 
$$
F = {9\over 2}\beta^2{N^2\over {4\pi r^3}} + 4\pi Sr^2 + {4\over 3}\pi r^3\left[B 
- {7\pi^2\over {180\beta^4}}\right]
\eqno{(47)}
$$
The scenario of soliton formation commences at temperatures greater than a critical temperature:
$$
{1\over \beta_c} = \left[180 B\over 7\pi^2\right]^{1/4} \approx 127~{\rm MeV ~~ for} ~~B^{1/4} = 100~{\rm MeV}
\eqno{(48)}
$$
The volume term in the expression for free energy is positive for temperatures lesser than $1/\beta_c$
and universe would be filled with massless fermions with the scalar
field locked at its value $\phi = 0$. At the critical temperature phase transition would commence with 
formation, expansion and percolation of bubbles of 
$\phi = \phi_o$ that would constrain the massless fermions to the old $\phi = 0$ phase domains. The  
temperature remains constant during the phase transition. The soliton bubbles contract till they are stabilized by the 
surface terms. We explore the stability under the assumption $\mu\beta_c <<1$. The expression for
energy becomes:
$$
E_c = \varrho_cv + 4\pi Sr^2 = \left[{9\over 2}\beta_c^2{N^2\over v} + {7\pi^2v\over {45\beta_c^4}}  \right]
+ 4\pi Sr^2
\eqno{(49)}
$$
The expression for the energy including gravitational effects follows from Eqn(36) by the 
replacement $\bar E \longrightarrow E_c$. The expression for free energy including gravitational effects is:
$$
F \simeq {3\over 2}\beta_c^2{N^2\over v} + 4\pi Sr^2 - {GE_c^2\over 
{r - m_c} + \sqrt{r^2 - 2m_cr}}; ~~ {\rm with}~~ m_c \equiv {GE_c\over c^2} \eqno{(50)}
$$
under a change of scale:
$$
r \equiv \left[{{9N^2\beta_c^2}\over {32\pi^2S}}\right]^{1/5} y\eqno{(51)}
$$
The expression for free energy becomes:
$$
F = \left[{{9N^2\beta_c^2}\over {8\pi}}\right]^{2/5}(4\pi S)^{3/5}\left[{1\over y^3} + y^2
- {{\epsilon_3y^5E^2(y)}\over {1 - \epsilon_4y^2E(y) + \sqrt{1 - 2\epsilon_4y^2E(y)}}}
\right]\eqno{(52)}  
$$
Here the function $E(y)$ and the constants in the above expression are 
$$
E(y) = 1 + {\epsilon_5\over 3y} + {\epsilon_5\over y^6}
$$
$$
\epsilon_3 \equiv 
G\left[{{9N^2\beta_c^2}\over {8\pi}}\right]^{3/5}(4\pi S)^{-8/5}\left[{16\pi\over 3}B\right]^2
$$
$$
\epsilon_4 \equiv 
G\left[{{9N^2\beta_c^2}\over {32\pi^2S}}\right]^{2/5}\left[{16\pi\over 3}B\right]
$$
$$
\epsilon_5 \equiv 
3\left[{{9N^2\beta_c^2}\over {8\pi}}\right]^{-1/5}(4\pi S)^{6/5}\left[{16\pi\over 3}B\right]^{-1} \eqno{(53)}
$$
For the usual choice for $S$ and $B$ used in hadron spectroscopy, we get:
$$
F = N^{4/5}\times 4.36\times 10^{-21}\left[{1\over y^3} + y^2
- {{\epsilon_3y^5E^2(y)}\over {1 - \epsilon_4y^2E(y) + \sqrt{1 - 2\epsilon_4y^2E(y)}}}
\right]\eqno{(54)}  
$$
with
$$
\epsilon_3 \simeq N^{6/5}\times 3.16\times 10^{-51};~~\epsilon_4 \equiv 5.1\times 10^{-49}N^{4/5}; ~~
\epsilon_5 \equiv  4.86 \times 10^8 N^{-2/5} \eqno{(54)}
$$
%-------------------figure for finite temp----------------------------
\begin{figure}[H]
%\centering
\mbox{\subfigure[]{\includegraphics[width=55mm, height=51mm]{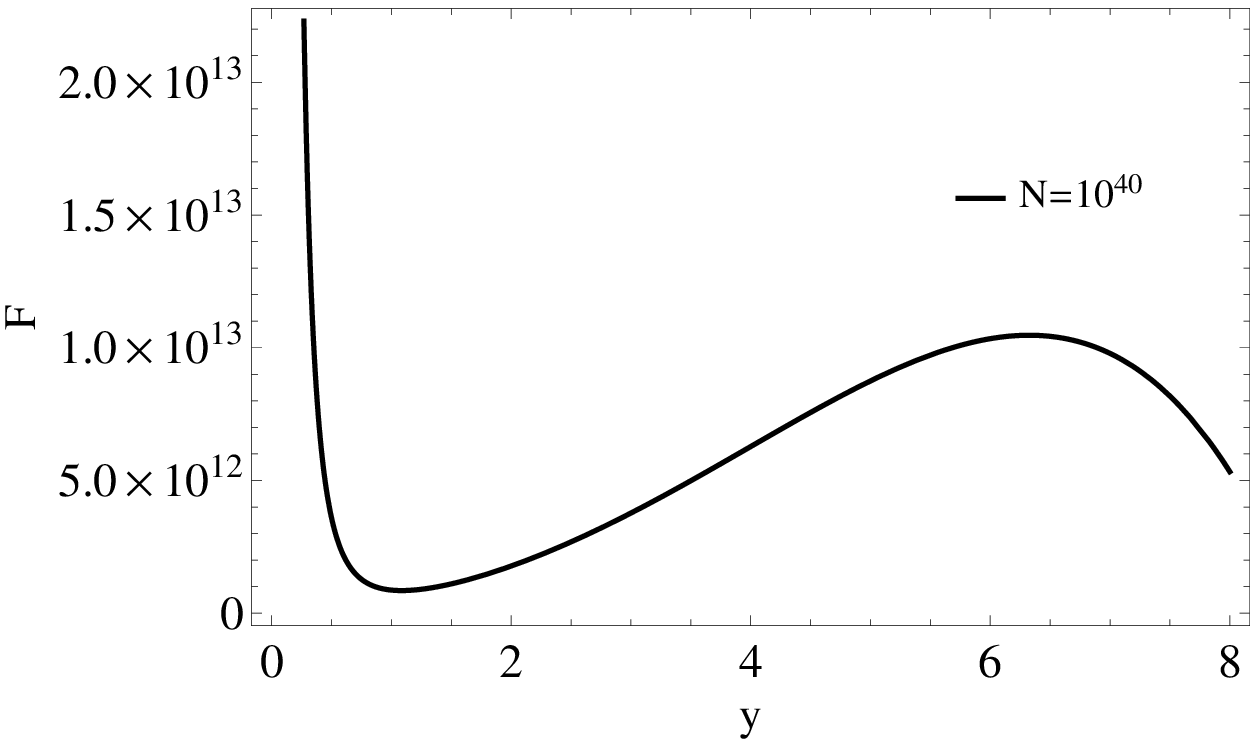}\label{fig3}}
\subfigure[]{\includegraphics[width=54mm, height=51mm]{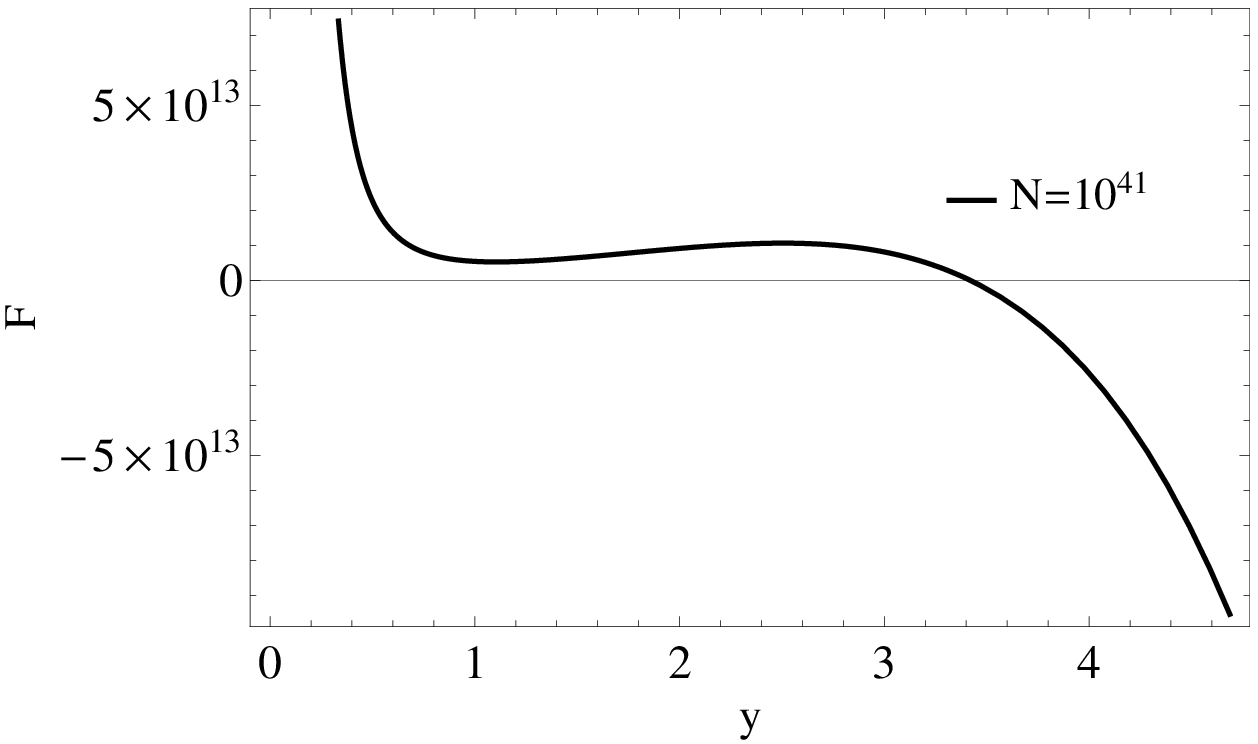}\label{fig4}}}
\caption{Variation of free energy at finite temperature with x.}
\end{figure}
The values for $S$ and $B$ in hadron spectroscopy are used in Figure(2) to find the limits: 
$N \geq 10^{41};~~ r_c \approx 1.15$~m$;~M_c \approx5.71*10^{17}$Kg. For neutrino balls, one could similarly 
determine these limits by chosing $B$ and $S$ accordingly.

\vspace{.5cm}
\section
{\bf Summary}
\vspace{.5cm}
Non-minimally coupled scalar field induced gravity theories
can support interesting non-topological soliton solutions. Our study would be useful in exploring the existence and formation of soliton stars that have a canonical effective gravitational constant in their exterior and vanishing gravitational "constant" in the interior. The existence of such configurations comes about on account of the "bag pressure" of a massless gas of fermions trapped in the interior of the soliton. The exact analytic expression of all thermodynamic parameters of the trapped gas, the total energy of the soliton, the entropy of the gas etc., can be written down. One of the prime motivations of such an endeavour was to explore how the expression for entropy for such configurations, for a given Schwarzschild exterior mass, would compare with the bounds conjectured for black hole entropy in standard general relativity. These aspects shall be reported elsewhere \cite{dlvart}.

\vspace{.5cm}
\section
{\bf Appendix:}
\vspace{.5cm}

Consider the Action given by eqns(1 \& 13). Requiring the action to be 
stationary under variations of the metric tensor and the fields 
$\phi, \psi$, gives the
equations of motion:
$$
U(\phi)[R^{\mu\nu} - {1\over 2}g^{\mu\nu}R] = -{1\over 2}[T_w^{\mu\nu} 
+ T_\phi^{\mu\nu} + T_{\phi,\psi}^{\mu\nu} + T_\psi^{\mu\nu}   
+ 2U(\phi)^{;\mu;\nu} 
- 2g^{\mu\nu}U(\phi)]^{;\lambda}_{;\lambda}] \eqno{(A.1)}
$$
$$
g^{\mu\nu}\phi_{;\mu;\nu} + {\partial V\over {\partial\phi}} 
- R{\partial U \over {\partial \phi}} + m_f'\bar\psi\psi  
= 0 \eqno{(A.2(a))}
$$
$$
\gamma^\mu D_\mu\psi + m_f\psi = 0\eqno{(A.2(b))}
$$
$$
D_\mu\bar\psi\gamma^\mu - m_f\bar\psi = 0\eqno{(A.2(c))}
$$
Here $T_w^{\mu\nu}$, $T_\psi^{\mu\nu}$and $T_{\phi,\psi}^{\mu\nu}$
are the energy momentum tensors constructed from
$L_w$ and $L_\psi + L_{\psi,\phi}$ respectively, and
$$
T_\phi^{\mu\nu} = \partial^\mu\phi\partial^\nu\phi 
- g^{\mu\nu}[{1\over 2}\partial^\lambda\phi\partial_\lambda\phi -V(\phi)]
\eqno{(A.3)}
$$
$L_w$ is  independent of $\phi$. To
examine this theory viz - a - viz the equivalence principle, we have
to explore conditions under which $T^{\mu\nu}_{w;\nu} = 0$.
Eqns.(2(a), 2(b)) show that the portion of Lagrangian 
$L_\psi + L_{\psi,\phi}$ is null on shell. The stress tensor is given by 
the following generalization of the familiar flat spacetime
expression:
$$
\Theta^\mu_\nu \equiv T_{\psi\nu}^\mu + T_{\psi,\phi,\nu}^\mu
= -{1\over 2}[\bar\psi\overleftarrow{D}_\nu\gamma^\mu\psi
- \bar\psi\gamma^\mu\overrightarrow{D}_\nu\psi]\eqno{(A.4)}
$$
When applied to any spinor or any ``spin - matrix'' such as the
Dirac matrices, one replaces the ordinary derivative by the spin -
covariant derivative \cite{pagels}. The covariant divergence of 
[A.4] is easily seen to reduce to:
$$
\Theta^\mu_{\nu;\mu} = m_f'\partial_\nu\phi\bar\psi\psi
\eqno{(A.5)}
$$
Thus there is a violation of equivalence principal as far as the 
Fermi field is concerned. However, in a region where the scalar
field gradient, $\partial_\mu\phi$, vanishes, the covariant divergence
of the fermion field stress tensor vanishes. For the rest of the
matter fields, the equivalence principal holds strictly, i.e.:
$T^{\mu\nu}_{w;\nu} = 0$. To see this, consider the covariant 
divergence of [A.1]. From the contracted 
Bianchi identity satisfied by the Einstein tensor, we get
$$
U(\phi)_{,\nu}[R^{\mu\nu} - {1\over 2}g^{\mu\nu}R] = 
-{1\over 2}[T^{\mu\nu}_{w;\nu} + t^{\mu\nu}_{;\nu}
+ \Theta^{\mu\nu}_{;\nu}] \eqno{(A.6)}
$$
with 
$$t^{\mu\nu} \equiv T_\phi^{\mu\nu} +  2U(\phi)^{;\mu;\nu} -
2g^{\mu\nu}U(\phi)^{;\lambda}_{;\lambda} \eqno{(A.7)}$$
Using the identity: 
$U(\phi)^{;\rho}R_{\rho\alpha} = U(\phi)^{;\lambda}_{;\lambda ;\alpha}
- U(\phi)^{;\lambda}_{;\alpha;\lambda}$ 
and the eqn(A.5), this reduces to
$$
-{1\over 2}U(\phi)^{,\mu}R = 
-{1\over 2}[T^{\mu\nu}_{w;\nu} + T^{\mu\nu}_{\phi;\nu}
+ (\partial^\mu m_f)\bar\psi\psi] 
$$
Finally, using the equation of motion for the scalar field [A.2a],
all the $\phi$ dependent terms cancel the left hand side -
giving the vanishing of the covariant divergence of the (w-) matter
stress energy tensor.
 
    One can find the expression for a conserved pseudo energy 
momentum tensor that would be conserved. To achieve this we proceed to 
express the vanishing covariant divergence of the matter stress energy 
tensor as:
$$
[\sqrt{-g}T^\nu_{w\mu}]_{,\nu} 
- {1\over 2}g_{\tau\beta,\mu} \sqrt{-g}T^{\tau\beta}_w = 0 \eqno{(A.8)}
$$
To cast the LHS of the above equation into a total ordinary divergence one 
has to seek a representation of the second quantity in terms of an
ordinary total divergence. This can be done as follows. 
First we make use of the equation of motion (A.1) 
to express the matter stress energy tensor in terms of the 
other fields and the metric -dependent quantities:
$\sqrt{-g}T_w^{\tau\beta} \equiv (-t^{\tau\beta} - \Theta^{\tau\beta} 
- 2U(\phi)G^{\tau\beta})\sqrt{-g}$. Next, we note 
that the right hand side of this expression is merely the variational
derivative of 
$$ 
J \equiv 2\int\sqrt{-g}d^4x[U(\phi)R + L_\phi 
+ L_\psi + L_{\psi,\phi}] \eqno{(A.9)}
$$
under variations of the metric tensor, with vanishing variation of the metric 
and its first derivative on the boundary of a (3+1) - dimensional manifold over 
which this  integral has been taken.
We consider the standard decomposition of 
$\sqrt{-g}R$ into a pure divergence term and a simple expression involving 
only the metric and its first derivatives:
$$
\sqrt{-g}R = {\rm A}
+ [\sqrt{-g}g^{\sigma\rho}\Gamma^\alpha_{\sigma\alpha}]_{,\rho}
-  [\sqrt{-g}g^{\sigma\rho}\Gamma^\alpha_{\sigma\rho}]_{,\alpha}\eqno{(A.10)}
$$
with 
$$
{\rm A} \equiv
\sqrt{-g}g^{\sigma\rho}[\Gamma^\alpha_{\sigma\rho}]\Gamma^\beta_{\alpha\beta}
- \Gamma^\alpha_{\beta\rho}]\Gamma^\beta_{\alpha\sigma}]\eqno{(A.11)}
$$
It follows that the functional derivative of $J$ with respect to the 
metric tensor is the same as that of  
$$
H \equiv \int d^4x [{\rm B} + \sqrt{-g}(L_\phi
+ L_\psi + L_{\psi,\phi})]\eqno{(A.12)}
$$
where
$$
{\rm B} \equiv [U{\rm A} 
- \sqrt{-g}g^{\sigma\rho}\Gamma^\alpha_{\sigma\alpha}U_{,\rho}
+ \sqrt{-g}g^{\sigma\rho}\Gamma^\alpha_{\sigma\rho}U_{,\alpha}]\eqno{(A.13)}
$$
In other words
$$
\sqrt{-g}UG_{\mu\nu} + 
\sqrt{-g}[U_{;\mu;\nu} - g_{\mu\nu}U^{;\alpha}_{;\alpha}]
+ {1\over 2}\sqrt{-g}T_{(\phi + \psi)\mu\nu}
$$
$$= {\partial\over \partial g^{\mu\nu}}[{\rm B} + \sqrt{-g}L_{\phi + \psi}]
- [{{\partial({\rm B} + \sqrt{-g}L_{\phi + \psi})}\over 
{\partial g^{\mu\nu}_{,\lambda}}}]_{,\lambda}\eqno{(A.14)}
$$
This is just a generalization of the standard procedure in General Relativity
\cite{bose}.
Defining $\hat {\rm B} \equiv {\rm B} + \sqrt{-g}L_{\phi + \psi}$, 
the expression for the 
ordinary derivative of $\hat {\rm B}$ and 
the field equation for the fields
$\phi, \psi$  easily enable us to express the 
second term in eqn(A.8) as a total
divergence. This gives
$$
[\sqrt{-g}T^\nu_{m\mu} - \hat {\rm B} \delta^\nu_\mu
+ {{\partial\hat{\rm B}}\over {\partial g^{\tau\beta}_{,\nu}}}g^{\tau\beta}_{,\mu}
+ {{\partial \hat{\rm B}}\over {\partial\phi_{,\nu}}}\phi_{,\mu}
+ {{\partial \hat{\rm B}}\over{\partial\psi_{,\nu}}}\psi_{,\mu}
+ \bar\psi_{,\mu}{{\partial \hat{\rm B}}\over{\partial\bar\psi_{,\nu}}}]_{,\nu}
= 0
\eqno{(A.15)}
$$
For $\nu = o$  the expression within the brackets integrated over
a spacelike hypersurface is thus invariant under time translations for 
a distribution of matter and the rest of the terms in  (A.15) having
a compact support over the surface. 
This is the expression for the pseudo energy momentum tensor that we seek.
The quantity
$$
P_\mu \equiv \int_\Sigma d\Sigma[\sqrt{-g}T^o_{w\mu} 
- \hat {\rm B} \delta^o_\mu
+ {{\partial\hat{\rm B}}\over {\partial g^{\tau\beta}_{,o}}}g^{\tau\beta}_{,\mu}
+ {{\partial \hat{\rm B}}\over {\partial\phi_{,o}}}\phi_{,\mu}
+ \bar\psi_{,\mu}{{\partial \hat{\rm B}}\over {\partial\bar\psi_{,o}}}
+ {{\partial \hat{\rm B}}\over {\partial\psi_{,o}}}\psi_{,\mu}
]\eqno{(A.16)}
$$
evaluated on a constant spacelike hypersurface $\Sigma$, is thus conserved. 
This may be viewed as the generalization of the energy momentum pseudo four vector
for the scalar - tensor theory described by eqns[1 \& 13]. The 
formalism presented here is general and can be used to determine
the energy momentum four vector for any Brans - Dicke theory in 
particular.
As in standard general relativity, $P_\mu$
is not a generally covariant four vector as ${\rm A}$ and ${\rm B}$ are not 
scalar densities. The intrinsic non - covariance of the energy momentum 
density of the gravitational field has its origin in the intimate 
connection between geometry and the gravitational field. Had the expression
been covariant, one could always have gone into a preferred [freely - falling] 
frame to ensure vanishing of an arbitrary localized gravitational field.

%        This expression for the energy is sufficient for the present article. 
%For a $U(\phi)$ that is well behaved [bounded], it is possible to
%reduce the energy as an integral over a two sphere. This is not 
%case for the present article [$U(\phi)$ has been chosen to diverge].
%[see Bose, Lohiya-- for the reduction]. 

The above form for the energy-momentum pseudo tensor for the 
generalized Brans-Dicke theory can also be obtained by considering a variation
of the coordinate system instead of the metric field. The analysis enables us 
to express the gravitational stress-energy pseudo tensor in a very compact 
form. Consider the variation of eqn[A-12]:
as a function of the metric, the scalar and the fermion fields, 
and their first derivatives: 
%\begin{eqnarray}
$$
\delta \hat {\rm B} = {\partial \hat {\rm B}\over \partial g^{\mu\nu}}\delta 
g^{\mu\nu} + {\partial\hat {\rm B}\over \partial g^{\mu\nu}_{,\lambda}}\delta 
g^{\mu\nu}_{,\lambda} + {\partial \hat {\rm B}\over \partial\phi}\delta\phi
+ {\partial\hat {\rm B}\over \partial\phi_{,\lambda}}\delta\phi_{,\lambda}
\nonumber 
$$
$$
+ {\partial \hat {\rm B}\over \partial\psi}\delta\psi
+ {\partial\hat {\rm B}\over \partial\psi_{,\lambda}}\delta\psi_{,\lambda}
+ {\partial \hat {\rm B}\over \partial\bar\psi}\delta\bar\psi
+ {\partial\hat {\rm B}\over \partial\bar\psi_{,\lambda}}\delta\bar\psi_{,\lambda}
\eqno{(A.17)}
$$
%\end{eqnarray}
Under an infinitesimal change of coordinates:
$$
\hat x^\alpha = x^\alpha + \epsilon\xi^\alpha \eqno{(A.18)}
$$
to the first order in $\epsilon$, we get the following variations:
$$
{\partial x^\alpha\over \partial\hat x^\lambda} = \delta^\alpha_\lambda 
- \epsilon {\partial \xi^\alpha\over \partial x^\lambda} + O(\epsilon^2)
$$
$$
\delta g^{\mu\nu} = \epsilon(\xi^\mu_{,\alpha}g^{\alpha\nu} +\xi^\nu_{,\alpha}
g^{\alpha\mu}) \ \ ,
$$
$$
\delta g^{\mu\nu}_{,\lambda} = \epsilon \left( g^{\tau\nu}_{,\lambda}
\xi^\mu_{,\tau} + g^{\mu\beta}_{,\lambda}\xi^\nu_{,\beta}
- g^{\mu\nu}_{,\alpha}\xi^\alpha_{,\lambda}
+ g^{\tau\nu}\xi^\mu_{,\tau ,\lambda}
+ g^{\tau\mu}\xi^\nu_{,\tau ,\lambda} \right)
$$
$$
\delta\sqrt{-g} = -\epsilon\sqrt{-g}\xi^\alpha_\alpha
$$
$$
\delta\phi = \delta\psi = \delta\bar\psi = 0
$$
$$
\delta(\phi_{,\lambda}) = -\epsilon\phi_{,\alpha}\xi^\alpha_{,\lambda}; ~~~
\delta(\psi_{,\lambda}) = -\epsilon\psi_{,\alpha}\xi^\alpha_{,\lambda}; ~~~
\delta(\bar\psi_{,\lambda}) = -\epsilon\bar\psi_{,\alpha}\xi^\alpha_{,\lambda}
$$
A restriction to linear transformations enables one to get an elegant form for
$\delta \hat {\rm B}$. The Christoffel symbols transform as tensors under such 
transformations and hence $\hat {\rm B}$ transforms as a scalar density. 
Thus 
$$
\delta \hat {\rm B} = {\hat {\rm B}\over \sqrt{-g}}\delta\sqrt{-g} = 
-\epsilon\xi^\alpha_{,\alpha}\hat {\rm B} \eqno{(A.19)}
$$
Substituting the above variations for an arbitrary linear 
coordinate transformation into eqn[A.17],  and comparing the expression with 
eqn[A.19], we obtain the identity:
$$
{\partial\hat{\rm B}\over \partial g^{\mu\nu}}g^{\alpha\nu}
+ {\partial\hat{\rm B}\over \partial g^{\mu\nu}_{,\lambda}}
g^{\alpha\nu}_{,\lambda} -{1\over 2} {\partial\hat{\rm B}\over \partial 
g^{\beta\nu}_{,\alpha}}g^{\beta\nu}_{,\mu}
- {1\over {2}}{\partial\hat {\rm B}\over \partial\phi_{,\alpha}}\phi_{,\mu}
- {1\over {2}}{\partial\hat {\rm B}\over \partial\psi_{,\alpha}}\psi_{,\mu}
- {1\over {2}}{\partial\hat {\rm B}\over \partial\bar\psi_{,\alpha}}\bar\psi_{,\mu}
= - {1\over 2}\hat {\rm B} g^\alpha_\mu \\
\eqno{(A.20)}
$$
Although the above identity was derived for variations under linear coordinate
transformations, one can verify that it holds quite generally. 
The use of this identity yields a simple expression for the variation of 
$\hat {\rm B}$ under the general transformation:
$$
\delta\hat {\rm B} = -\epsilon\hat {\rm B}\xi^\alpha_{,\alpha} + 
2\epsilon{\partial\hat {\rm B}\over \partial g^{\mu\nu}_{,\lambda}}
\xi^\mu_{,\tau ,\lambda}g^{\tau\nu} \eqno{(A.21)}
$$
Under conditions where $\xi$ and its derivatives are taken to vanish on the 
boundary, the variation of the metric tensor and its derivatives also vanish 
there. Under such boundary conditions, H has a vanishing variation, i.e.,
$$
\delta H = \int_\Sigma\delta \left({\hat {\rm B} \over \sqrt{-g}} \right)
\sqrt{-g}\> d^{(D+1)}x = 0 \eqno{(A.22)}
$$
which, using the above identities, reduces to
$$
\delta H = 
2\epsilon\int_\Sigma {\partial\hat {\rm B}\over \partial 
g^{\mu\nu}_{,\lambda}}\xi^\mu_{,\tau ,\lambda}g^{\tau\nu}d^{(D+1)}x = 0 \eqno{(A.23)}
$$
This expression may be integrated by parts twice. Since $\delta H$ vanishes 
for arbitrary $\xi^\mu$, we obtain the following divergence law:
$$
\left({\partial\hat {\rm B}\over \partial g^{\mu\nu}_{,\lambda}}
g^{\tau\nu}\right)_{,\tau ,\lambda} = 0 \eqno{(A.24)}
$$
Thus 
$$
\sqrt{-g}F^\tau_\mu \equiv \left({\partial\hat {\rm B}\over \partial 
g^{\mu\nu}_{,\lambda}} g^{\tau\nu} \right)_{,\lambda} \eqno{(A.25)}
$$
defines a conserved quantity. Using the identity (A.20) and the field 
equation (A.14) gives:
$$
\sqrt{-g}F^\tau_\mu = - {1\over 2}\sqrt{-g}T^{(m)\tau}_\mu 
-{1\over 2}\hat {\rm B}g^\tau_\mu 
+ {1\over 2}({\partial\hat {\rm B}\over \partial 
g^{\beta\nu}_{,\tau}}g^{\beta\nu})_{,\mu}
$$
$$ + {1\over 2}{\partial\hat {\rm B}
\over \partial\phi_{,\tau}}\phi_{,\mu}  + {1\over 2}{\partial\hat {\rm B}
\over \partial\psi_{,\tau}}\psi_{,\mu} 
+ {1\over 2}\bar\psi_{,\mu}{\partial\hat {\rm B}
\over \partial\bar\psi_{,\tau}} \eqno{(A.26)}
$$
which is just the expression that we had obtained for the stress energy 
pseudo tensor by the variation of the metric tensor earlier. The expression 
(A.24) for a vanishing ordinary divergence implies that 
$$
P_\mu \equiv - \int_V \left({\partial\hat {\rm B}\over 
  \partial g^{\mu\nu}_{,\lambda}}g^{o \nu} \right)_{,\lambda}dV \eqno{(A.27)}
$$
is a conserved quantity if $V$ is the entire space at a given time. In the 
special case of a time independent metric, there is no sum over $\lambda = 0$,
and Gauss's theorem in 3 - dimensions 
gives the energy momentum as surface integral over a 2 -dimensional 
surface:
$$
P_\mu = - \int_\Sigma\left({\partial\hat {\rm B}\over \partial 
     g^{\mu\nu}_{,j}} g^{o \nu}\right)d\Sigma_j \eqno{(A.28)}
$$
This gives the interesting result that in the generalized Brans-Dicke theory, 
the generalized energy-momentum in a 3-dimensional volume is determined 
by the metric-tensor and its derivatives on surface the volume, 
the details of the field inside the volume being irrelevant.

For a spherically symmetric solution that has the scalar field locked 
to the minimum of the potential $V(\phi)$ and is empty outside a finite compact 
region, the solution to the field equation is simply the Schwarzschild solution
with the geometric mass parameter $m$.
From the expression of the Schwarzschild solution in isotropic coordinates (that
are used as the metric in these coordinates approaches the Lorentz metric at 
infinity), the asymptotic value of the energy is simply the total mass
of the system:
$$
P_o = M = mc^2 \times 8\pi U(\phi_o) \equiv {mc^2\over {2G}} \eqno{(A.29)}
$$
For the scalar field locked at a minimum of the potential $V(\phi)$ outside a 
spherically symmetric domain, the exterior metric in that domain, in isotropic
coordinates is simply:
$$
ds^2 = {(1 - m/2\rho)^2\over {(1 + m/2\rho)^2}}dt^2 - (1 + {m\over 2\rho})^4
(d\rho^2 + \rho^2(d\theta^2 + sin^2\theta d\varphi^2) \eqno{(A.30)}
$$        
The total conserved (pseudo) energy inside a sphere of radius $\rho = \rho_o$, can be 
easily determined to be:
$$
P_o = M\left(1 - {M\over\ {32\pi U(\phi_o) c^2 \rho_o}}\right) 
= M\left(1 - {MG\over\ {2c^2 \rho_o}}\right) \eqno{(A.31)}
$$
This is an exact expression for the energy of a shell of radius $\rho = \rho_o$ with an 
asymptotic mass $M$. 
If we now assume that the non-minimal coupling function makes a sharp transition from
its exterior value $U(\phi_o)$ to a large value $U(0) \longrightarrow \infty$ across 
a thin wall, the above analysis establishes that the energy inside the sphere 
of radius $R_o$ given by eqn[A.31] is simply rearranged in the flat spacetime inside
the shell where the interior metric is simply:
$$
ds^2 = dt_{in}^2 - dr^2 - r^2(d\theta^2 + sin^2\theta d\varphi^2) \eqno{(A.32)}
$$
with 
$$
dt_{in} \equiv \left({(1 - m/2\rho_o)\over {(1 + m/2\rho_o)}}\right)dt; ~~~~~ 
r \equiv (1 + {m\over 2\rho_o})^2\rho
$$
The expression for the volume, surface area and the energy of a shell of radius 
$r = r_o \Longleftrightarrow \rho = \rho_o$ is:
$$
v = {4\over 3}\pi r_o^3; ~~~ A = 4\pi r_o^2:
$$
$$
P_o = M\left(1 - {MG\over {2c^2 \rho_o}}\right)
= M\left(1 - {MG\over {c^2(r_o - m + \sqrt{r_o^2 - 2mr_o}}}\right)\eqno{(A.33)}
$$
The stability properties of such a configuration, in which the asymptotic energy $M$ is
distributed over a volume term with $\phi = \phi_o$ in the interior; a surface term;  
and the energy of a fermion gas, is considered in section 3.

\vfil
\eject 

\vskip 0.5 cm
{\bf Acknowledgments}: We are grateful to Ashok Goyal for essential clarifications. DL thanks DAMTP, University of Cambridge for assistance. VG thanks CSIR (India) for financial support through grant number 09/045/(0933)/2010 EMR-I.

\vskip 1cm

\bibliography{plain}

\vfil
\eject

\end{document}